\documentclass[twocolumn,showpacs,preprintnumbers,amsmath,amssymb,aps,10pt,prb]{revtex4-1}
\usepackage{graphicx}
\usepackage{verbatim} 
\usepackage[dvips]{color} 

\begin{document}

\title{Temperature dependence of hole spin coherence in (In,Ga)As quantum dots\\ measured by mode-locking and echo techniques}

\author{S. Varwig$^1$, A. Ren\'e$^1$, A. Greilich$^1$, D.~R.
Yakovlev$^{1,2}$, D. Reuter$^3$, A.~D. Wieck$^3$ and M. Bayer$^1$}

\affiliation{$^1$ Experimentelle Physik 2, Technische Universit\"at
Dortmund, 44221 Dortmund, Germany}

\affiliation{$^2$ Ioffe Physical-Technical Institute, Russian
Academy of Sciences, 194021 St. Petersburg, Russia}

\affiliation{$^3$ Angewandte Festk\"orperphysik, Ruhr-Universit\"at
Bochum, 44780 Bochum, Germany}

\begin{abstract}
The temperature dependence of the coherence time of hole spins
confined in self-assembled (In,Ga)As/GaAs quantum dots is studied
by spin mode-locking and spin echo techniques. Coherence times
limited to about a $\mu$s are measured for temperatures below 8\,K. For higher
temperatures a fast drop occurs down to a few ns over a 10\,K range.
The hole-nuclear hyperfine interaction appears too weak to account for these
limitations. We suggest that spin-orbit related interactions
are the decisive sources for hole spin decoherence.
\end{abstract}


\maketitle

During recent years the spins of holes confined in III-V
semiconductor quantum dots (QDs) have attracted considerable interest.
This interest is related to the hole spin's hyperfine coupling to nuclear spins, which has been found to be non-negligible, in contrast to original suggestions based on the vanishing contact interaction, but still considerably reduced by an order of magnitude
as compared to electron spins.\cite{Fallahi2010,Chekhovich2011,Varwig2012} 
For the electrons the hyperfine
interaction has been identified as the source of spin decoherence at
cryogenic temperatures:\cite{Merkulov2002,Khaetskii2002} the transverse spin relaxation time is on the order of microseconds at these conditions,\cite{Petta2005,Greilich2006} well below the longitudinal spin relaxation times of up to milliseconds in
magnetic fields large enough for sizable two-level splittings that
might be of use in quantum information.

From their reduced hyperfine coupling longer coherence times may be
expected for the hole spins.\cite{Fischer2010} However, recent studies have
demonstrated that the hole spin coherence time $T_2$ is {\sl not} elongated, but rather comparable to that of the electron with values on the
order of a $\mu$s.\cite{Greve2011,Varwig2012,Fras2012} The origin of this
behavior is not yet understood: spin-orbit interaction as another
potential spin relaxation mechanism involving phonons may be more
important for holes than for electrons, but at liquid helium
temperatures it should be suppressed in quantum dots with a
discrete energy level structure. To obtain more insight into the
problem, it might be helpful to study the temperature dependence of
the hole spin coherence time.

This is the problem that we address here by exploiting the recently
demonstrated hole spin mode-locking (SML) in consequence of periodic
pulsed excitation of the ground state transition by circularly
polarized laser light.\cite{Varwig2012} By monitoring the SML signal
amplitude in dependence of the laser pulse separation at various temperatures we find that the hole spin coherence time remains constant in the $\mu$s range
only up to about 8\,K, but then drops quickly down to nanoseconds at
20\,K, approaching the lifetime of optically excited electron-hole
pairs. This result is confirmed by detecting the temperature dependence
of optically induced hole spin echoes. From this we conclude that spin-orbit interactions play
the decisive role, even though details of the mechanism need further
elaboration.

\begin{figure}[t]
\includegraphics[width=\columnwidth]{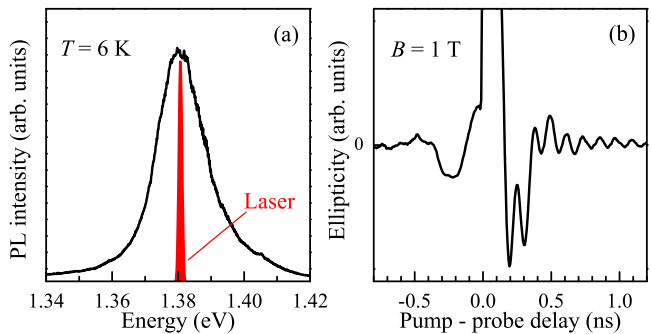}
\caption{(Color online) (a) Photoluminescence spectrum of the
(In,Ga)As/GaAs QD ensemble, measured at $T=6$\,K. The red-colored line
shows the laser spectrum with a FWHM of 1\,meV which is in resonance
with the QD ground state emission. (b) Pump-probe ellipticity
measurement at $B=1$\,T and $T=6$\,K. The zero delay peak influenced
by scattered laser light is cut off for better visibility of the
spin precession oscillations.} \label{fig:fig1}
\end{figure}

The sample under study was grown by molecular-beam
epitaxy on a (001) GaAs substrate and contains ten layers
of (In,Ga)As dots, separated by 100\,nm GaAs barriers. The
QD density per layer is about $10^{10}$\,cm$^{-2}$. The dots are
nominally undoped, but it has been found in  earlier studies that about half of them are singly positively charged due to residual carbon
impurities.\cite{Varwig2012} The sample was annealed for 30\,s at a temperature of
960\,$^\circ$C, leading to a band gap increase such that resonant
excitation by a Ti:Sapphire laser is possible. The photoluminescence
(PL) spectrum in Fig.~\ref{fig:fig1}(a) shows the ground state emission with a maximum at 1.38\,eV and a full width at half maximum (FWHM) of about 20\,meV.

To study the QD spin dynamics we use a degenerate pump-probe setup with the
laser energies tuned to the PL maximum. The optical axis ($z$-axis)
is chosen parallel to the sample growth direction.  Spin
polarization of the QD hole is generated by a periodic train of circularly polarized
pump pulses exciting the transition from the resident hole to the
positively charged exciton. The spin polarization along the $z$-axis
is monitored by measuring the ellipticity of an originally linearly
polarized probe beam after transmission through the sample, which is mounted in a cryostat allowing variable temperatures $T$ down to 2\,K and magnetic fields $B$ up to 7\,T. Application of such an external magnetic field along the $x$-axis (Voigt geometry) leads to precessions of the hole spins about this axis.

Pump and probe pulses are taken from a Ti:Sapphire laser operating  at a
repetition frequency of 75.75\,MHz, corresponding to a pulse separation
of $T_R = 13.2$\,ns that can be extended by a pulse picker. The
pulses with a duration of 1.5\,ps have a spectral width of 1\,meV so that an ensemble of about $10^5$ QDs is addressed. The pump pulse intensity is adjusted to a pulse area of $\pi$ and the probe pulse intensity is taken five times weaker. By varying the delay between pump and probe pulses, the
temporal evolution of the spin polarization is measured, as shown in
Fig.~\ref{fig:fig1}(b).

After pump incidence at zero delay one sees damped oscillations with
contributions from resident and photocreated electron and hole spins. These contributions are distinguishable by their precession frequencies due to different $g$-factors, namely $|g_e| = 0.58$ for electrons and $|g_h| = 0.14$ for holes. The damping of
the signal arises from spin dephasing due to $g$-factor variations in the ensemble. In Ref.~\onlinecite{Yugova2007} we have shown that for the in-plane hole $g$-factor these variations are comparable in magnitude
to the absolute $g$-factor value. Therefore the hole spin dephasing time $T_{2,h}^* \sim 0.25$\,ns is quite fast, compared with the electron spin dephasing time of $T_{2,e}^* \sim 1.03$\,ns.

At negative delays, before pump pulse arrival, a rephasing of the
resident hole spins due to the spin mode-locking (SML) effect is visible. This
rephasing arises from spins whose precession about the magnetic
field becomes synchronized with the laser pulse repetition rate, as
expressed by the phase synchronization condition (PSC):\cite{Greilich2006}
\begin{equation}
\omega = \frac{|g_{h}| \mu_B B}{\hbar}= N \dfrac{2\pi}{T_R},
\label{PSC}
\end{equation}
where $\omega$ is the Larmor precession frequency determined by the hole
$g$-factor $g_h$ along the magnetic field $B$, $\mu_B$ is the Bohr magneton, and $N$ is a positive integer. Note, however, that the hole spin mode-locking
amplitude is weaker by a factor of three compared to the amplitude at positive delays as a result of the rather weak hyperfine interaction: the coupling to the nuclei, if efficient, would drive modes which do not initially fulfill the PSC into spin mode-locking. This nuclear frequency focusing would occur by building up a nuclear field of proper strength that adds to the external magnetic field. While being efficient for electrons, for which
after sufficiently long pumping all optically excited ones can
contribute to mode-locking,\cite{Greilich2007} the hyperfine interaction for holes is too weak  to induce such nuclear frequency focusing.\cite{Varwig2012} As a consequence the
negative delay signal is considerably weaker than the one at positive delays, where
the signal strength is determined by the entirety of excited dots.

For spin echo experiments an additional pulse, termed the control
pulse, is introduced into the excitation scheme. This pulse is taken
from a second Ti:Sapphire laser system synchronized with the first
one with a 100\,fs accuracy, but independently tunable in photon energy. Pulse duration and spectral width are equal for the two lasers. The second laser is
used to rotate the hole spins about the optical $z$-axis, in analogy
to experiments on electron spin rotations in (In,Ga)As/GaAs QDs.\cite{greilich2009NP} The control
pulse intensity is adjusted to a pulse area of $2 \pi$ in order not to
populate trion states, but remain in the spin subspace of resident holes.

\begin{figure}[t]
\includegraphics[width=\columnwidth]{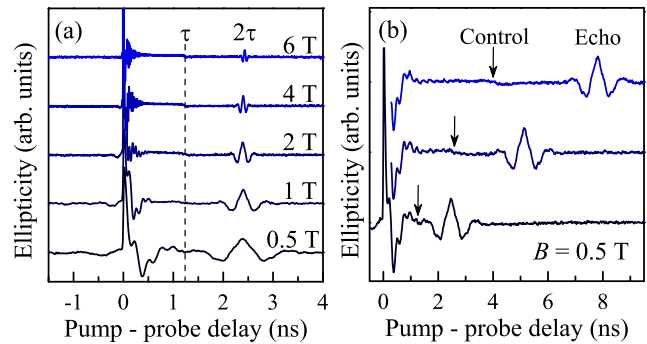}
\caption{(Color online) (a) Hole spin echoes at different magnetic
fields, $\tau=1.2$\,ns, $T=6$\,K. Curves are shifted vertically for clarity. (b)
Hole spin echoes for different control pulse delays $\tau=1.2, 2.6$ and 3.9\,ns. The echo
appearance time $2\tau$ shifts in accordance with the control pulse
arrival time $\tau$. $B=0.5$\,T and $T=6$\,K.} \label{fig:fig2}
\end{figure}

By adjusting the control photon energy to the pump photon energy each control pulse rotates the resident hole spins in those QDs excited by the pump pulse by an angle
of $\pi$. Figure~\ref{fig:fig2}(a) shows spin rotation measurements
at different magnetic fields varied from 0.5 to 6\,T. The control pulse
hits the sample at a time delay $\tau = 1.2$\,ns relative to the pump pulse
arrival and at a time $2\tau = 2.4$\,ns a hole spin echo appears. It arises from
the $180^{\circ}$ rotation of the hole spin ensemble at delay $\tau$
by which the dephasing occurring between pump and control is
inverted and the spins reconvene. The temporal sequence is
confirmed when the control pulse delay $\tau$ is varied, as seen in Fig.~\ref{fig:fig2}(b). The time between echo formation and pump arrival is twice the time between control
and pump.

\begin{figure}[t]
\includegraphics[width=\columnwidth]{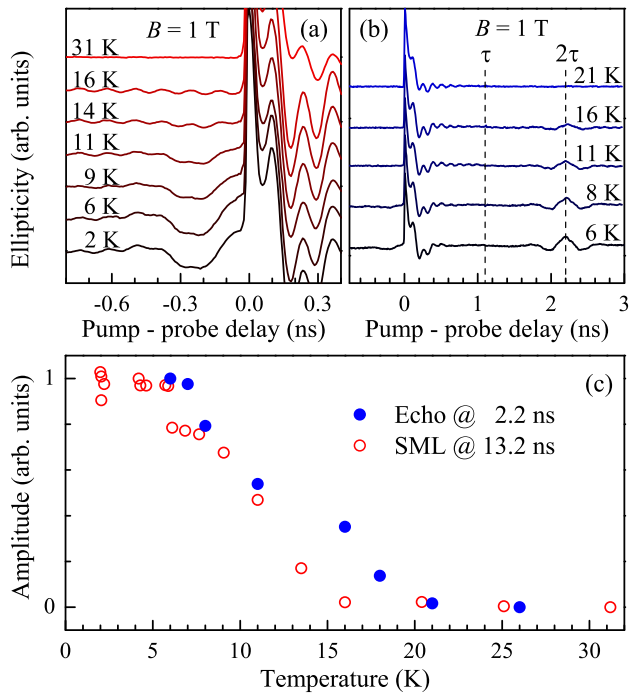}
\caption{(Color online) (a) Time-resolved ellipticity measurements with
focus on negative delay times for different temperatures, $B=1$\,T and $T_R = 13.2$\,ns. Fits to the hole spin oscillations (see text) provide the SML amplitudes (red open circles) in panel (c). (b) Hole spin echoes recorded at different temperatures for $B=1$\,T, $T_R=13.2$\,ns and $\tau=1.1$\,ns. From the echo oscillations we deduce the blue data points (solid circles) in panel (c). (c) Temperature dependence of the normalized hole spin echo (blue solid circles) and SML (red open circles) amplitudes.} \label{fig:fig3}
\end{figure}

The two effects, spin mode-locking and spin echo,
are exploited to determine the temperature dependence of the hole spin
coherence time. First we turn to the spin mode-locking signal in pure pump-probe studies without control pulses. Corresponding ellipticity traces are shown in
Fig.~\ref{fig:fig3}(a) with focus on the mode-locking signal at
negative delay times, recorded for different temperatures. The pump
pulse separation was 13.2\,ns. Two contributions with different
frequencies are seen in the signal; the one with lower frequency
is related to the hole spin precession of interest here. Because of
the strong damping, even at the lowest temperatures only one full
oscillation is seen. The higher frequency contribution can be
assigned to the electron spin.

Clear hole spin mode-locking (SML) can be seen in Fig.~\ref{fig:fig3}(a) up to temperatures of 15\,K. From these observations one identifies two temperature ranges, for which we have to employ different methods for extracting the hole spin coherence time $T_2$. At the lower end of the studied temperatures ($T \lesssim 10$\,K) $T_2$ is longer than the pulse repetition period $T_R$ of 13.2 ns, such that SML can occur.
The appropriate method then is to increase $T_R$ until SML becomes weaker. From the variation of the SML amplitude with $T_R$, the spin coherence time can
be extracted.\cite{Greilich2006,Varwig2012} Additionally, if $T_2$ is shorter than $T_R$, as apparently is the case for elevated temperatures, we can estimate
the spin coherence time from the temperature dependence of the
SML amplitude and the spin-echo amplitude (see below).

\begin{figure}[t]
\includegraphics[width=\columnwidth]{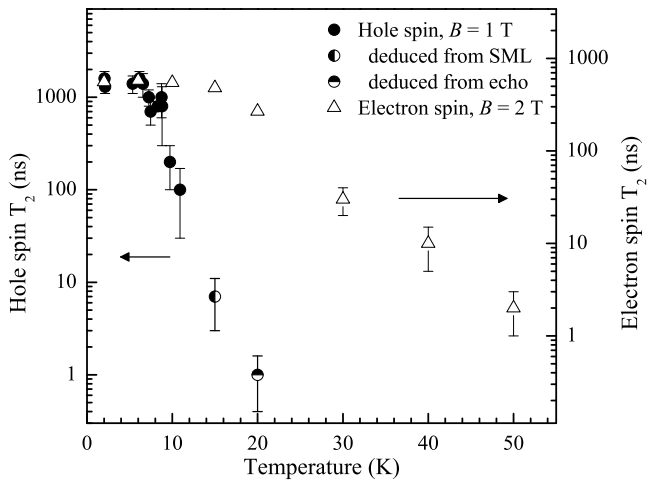}
\caption{Temperature dependence of the
coherence time $T_2$ for hole spins (circles, left scale) and electron
spins (triangles, right scale). The filled circles are measured by mode-locking experiments with various pulse separations at different temperatures (see text); $B=1$\,T. The vertically halved circle is deduced from
the temperature dependence of the mode-locked spin amplitude with $T_R = 13.2$\,ns, see Figs.~\ref{fig:fig3}(a) and \ref{fig:fig3}(c). The horizontally halved circle is deduced from the temperature dependence of the hole spin echo,
see Figs.~\ref{fig:fig3}(b) and \ref{fig:fig3}(c). The electron data is taken from
Ref.~\onlinecite{Hernandez2008} being measured at $B=2$\,T.
They are valid for comparison as from 1\,T up to 3\,T no magnetic field dependence was
observed. } \label{fig:fig4}
\end{figure}

Let us first concentrate on the low-temperature regime. To vary $T_R$ we
implemented a pulse picker in the setup, which reduces the pulse
repetition rate by letting only particular pulses of the original
train pass while dumping the others. In that way we increased the
time between two subsequent pump pulses incrementally from
$T_R=132$ up to 660\,ns and extracted the spin mode-locking amplitude $A_{\text{SML}}$ using a cosine fit with a Gaussian damping
function. One can then determine the dependence of this amplitude on $T_R$ and obtain the coherence time
$T_2$ using the following relation: 
\begin{equation}
A_{\text{SML}}(T_R) \propto \exp\left[-\left(2+\frac{1}{2\sqrt{3}+3}\right)\frac{T_R}{T_2}\right],
\label{T2}
\end{equation}
which can be found in the SOM of Ref.~\onlinecite{Greilich2006}. The results for $T_2$ are shown in Fig.~\ref{fig:fig4} by filled circles. The hole
spin coherence time is constant at slightly more than 1\,$\mu$s up to
almost 6-7\,K, but then drops to about 100\,ns at 10\,K. A further temperature increase makes the SML disappear already for a pulse separation of
132\,ns, which was the shortest that could technically be achieved with the pulse picker.

In the elevated temperature range $T > 10$\,K the other methods need to be
applied, directly exploiting the temperature dependencies of
spectroscopic quantities such as the SML amplitude. While this
procedure is straightforward, only estimates for $T_2$ can be
obtained in this way. Using again a Gaussian damped cosine fit, the amplitudes of the mode-locked resident hole spin
polarization in Fig.~\ref{fig:fig3}(a) can be extracted. These amplitudes are
plotted in Fig.~\ref{fig:fig3}(c). The drop of the amplitude between 2\,K and 15\,K occurs because the spin coherence time has become comparable or shorter than the pulse separation of 13.2\,ns. From the SML amplitudes before and after the drop, as seen
in Fig.~\ref{fig:fig3}(c), one can estimate that the hole spin coherence time has to be around 7\,ns at 15\,K. The corresponding data point is inscribed in Fig.~\ref{fig:fig4} by the vertically halved circle.

Another data point can be obtained from the temperature dependence of
the hole spin echo signal. In these studies the control
pulse hits the dephased hole spin ensemble at 1.1\,ns delay,
leading to an echo at 2.2\,ns, as seen in Fig.~\ref{fig:fig3}(b). 
Similar to the SML signal, the echo amplitude is constant at low temperatures, but starts to decrease strongly at around 10\,K and vanishes around 20\,K. For
analysis, the signal around the echo is also fitted by a cosine
function with a Gaussian amplitude envelope. This amplitude is
plotted in Fig.~\ref{fig:fig3}(c), closely resembling the
corresponding dependence for the SML signal. From this dependence we
get a data point of $T_2 = 1$\,ns at $T= 20$\,K.

Figure~4 shows the temperature dependence of the hole spin coherence
time and, for comparison, the electron spin coherence time in a similar quantum dot sample,
published in Ref.~\onlinecite{Hernandez2008}.
Electron and hole spin coherence times are constant at low
temperatures, but then quickly decrease into the ns-range at moderate temperatures. While for the holes this decrease is very abrupt and starts at 8\,K, for the electrons $T_2$ is constant in the $\mu$s-range up to slightly more than 15\,K before the drop occurs over a
30\,K range. For the electrons the drop was associated to elastic
scattering due to phonon-mediated fluctuations of the hyperfine
interaction.\cite{Hernandez2008}

Both the limitation of the hole spin coherence below a temperature of 8\,K and the
drop for higher temperatures need to be explained. Carrier spin
coherence in quantum dots has been considered in a
few works in which either hyperfine interactions or spin-orbit
related interactions were addressed as sources for decoherence. Let
us first discuss the works involving the nuclear bath: in
the original work by Fischer and Loss \cite{Fischer2008} it was shown that while
the contact interaction is not relevant for holes, other
contributions, the dipole-dipole interaction and the orbital angular
momentum interaction with nuclear spins, can become important and
may be even of comparable strength as the electron contact hyperfine
interaction. Indications to that end were reported in Refs.~\onlinecite{Testelin2009}.

It was recently shown, however, that the hyperfine interaction is at
least an order of magnitude weaker for the holes then it is for electrons in self-assembled (In,Ga)As QDs.\cite{Fallahi2010,Chekhovich2011,Greve2011,Wang2012,Varwig2012,Fras2012,Eble2009} Based on these results it is unlikely that
the interaction with the nuclei leads to the $\mu$s-limit for the hole
spin coherence. As mentioned, for the temperature dependence of the
electron spin coherence a model was also developed based on the
spectral diffusion of the nuclear-spin distribution due to
excitation of acoustic phonons.\cite{Semenov2007,Hernandez2008}  While it works well for electrons, due to the reduced interaction strength this
mechanism also seems unlikely to explain the temperature dependence for holes.

If we neglect the hole spin interaction with the nuclei, the spin-orbit
interaction must account for the observed behavior of the hole spin
coherence time $T_2$. The hole might be excited to an excited orbital state  having a different $g$-factor without spin-flip, resulting in a different
precession frequency. After relaxation, this would lead to a phase change
of the spin precession.\cite{Semenov2007} In addition, the impact of
anharmonic phonons due to lattice impurities, defects, etc. has been
emphasized. Interactions with them can lead to a phase change in the
coherent spin dynamics.\cite{Semenov2007} Both mechanisms formulated for
electrons are obviously also relevant for holes,
potentially even more prominently as the spin-orbit interaction is
stronger in the valence band.

From the splitting of about 20\,meV between the first excited state and the ground state,
obtained from high excitation photoluminescence studies, we estimate a splitting of the corresponding valence band states of at least 5\,meV,\cite{Zibika2005} whereas the temperature of 8\,K at which the sharp drop of $T_2$ sets in is considerably less than 1\,meV. The
excitation to excited hole states can therefore be ruled out under these
conditions. Crystal defects in self-assembled QDs have been shown to
be strongly suppressed, while the alloying (corresponding to some
extent to implementation of impurities disturbing the crystal
periodicity) broadens the phonon spectrum. In any case phonon activated mechanisms require thermal excitation to higher states, so most likely neither can be responsible for the low temperature limitation.

Consequently, we suggest here a different mechanism. For quantum dots,
the interaction of carriers with acoustic phonons has been shown to
have important consequences as it limits the quantum mechanical
coherence of confined carriers. For example, due to it, the
coherent exciton polarization drops over a timescale of a few picoseconds after
pulsed carrier excitation.\cite{Kurtze2009} In the spectral domain, this scattering
results in broad spectral flanks of the zero-phonon spectral line of
the QD exciton transition.\cite{Borri2001,Vagov2002} The underlying
mechanism can be understood as follows: the carriers lead to the
formation of a quasi-stable polaron, a bound state of the injected
charges and an associated phonon population, by which the lattice
becomes distorted. As a result of this distortion, a coherent phonon
wave packet is emitted from the QDs, escaping on timescales of
ps into the embedding material.

Another example for the importance of these phonon sidebands is the light emission from
QDs into a laser resonator mode, as not only the dots in
resonance with this mode contribute to the emission, but also
off-resonant quantum dots in which the carriers can
recombine and the photon is funneled into the laser mode with
simultaneous phonon scattering. In that way the laser threshold is
reduced and the laser output increased.\cite{Kaniber2008,Press2007,Hennessy2007,Winger2009,Ates2009}

Due to the efficiency of this coupling we suggest that, through the
phonon sidebands reflecting the polaronic lattice distortion, elastic
spin scattering becomes possible, leading to a destruction of the
phase coherence. Double phonon processes scattering out of one level
and back into the same level may become efficient because the
crystal offers a continuum of phonon states, enabling multiple
scattering options. This scattering may explain the restriction of
the hole spin coherence time to microseconds below 8\,K and might also
explain the speed-up of spin relaxation at higher temperatures. However, still more detailed studies need to be done.

In summary, we have studied the temperature dependence of the hole
spin coherence time $T_2$ by making use of the spin mode-locking effect and all-optically created hole spin echoes. The sharp drop of $T_2$ from $\mu$s down to ns
below 20\,K suggests that mechanisms considered so far based on the
hyperfine coupling and the spin-orbit interaction are not adequate
for explanation and that further considerations may be required. We
suggested here an alternative mechanism relying on elastic scattering
exploiting the broad phonon sidebands. Prospectives for the further
coherent manipulation of hole spins include multi-echo techniques, which might be used in dynamic decoupling schemes by which the spin coherence time
might be extended, like in NMR experiments.\cite{Bluhm2010}

This work was supported by the Deutsche Forschungsgemeinschaft and the BMBF project QuaHL-Rep.

\end{document}